\documentclass[aps,prl,nobibnotes,nofootinbib,showpacs,reprint]{revtex4-1}

\usepackage{graphicx}
\usepackage{amsmath} 	

\newcommand{\Fig}[1]{figure~\ref{#1}}
\newcommand{\Eq}[1]{equation~(\ref{#1})}
\newcommand{\vect}[1]{\mathbf{#1}}
\newcommand{\eps}{\epsilon_{0}}

\newcommand{\net}{\textrm{net}}
\newcommand{\eff}{\textrm{eff}}
\newcommand{\infint}{\int_{-\infty}^{+\infty}}

\UseRawInputEncoding
\begin{document}

\title{A Low-Power Optical Electron Switch}

\author{Wayne~Cheng-Wei~Huang}
\email{email: waynehuang1984@gmail.com}

\author{Roger~Bach}
\email{email: roger.bach@gmail.com}

\author{Peter~Beierle}
\email{email: pjbeierle@gmail.com}

\author{Herman~Batelaan}
\email{email: hbatelaan2@unl.edu}

\affiliation{Department of Physics and Astronomy, University of Nebraska-Lincoln, Lincoln, Nebraska 68588, USA}

\begin{abstract}
	An electron beam is deflected when it passes over a silicon nitride surface, if the surface is illuminated by a low-power continuous-wave diode laser. A deflection angle of up-to $1.2 \,\textrm{mrad}$ is achieved for an electron beam of $29 \,\mu\textrm{rad}$ divergence. A mechanical beam-stop is used to demonstrate that the effect can act as an optical electron switch with a rise and fall time of $6  \,\mu\textrm{s}$. Such a switch provides an alternative means to control electron beams, which may be useful in electron lithography and microscopy.

\end{abstract}

\maketitle

The motion of electron beams is controlled in technologies such as electron lithography, microscopy, and diffractometry, in which the use of electric and magnetic fields to focus and steer beams are proven techniques. The control of electron motion with laser fields is also possible with the ponderomotive potential \cite{Boot, Sciaini}. In principle, such a technique offers the interesting possibility that no electrical components or other hardware needs to be placed in the vicinity of the electron beam. In addition, using the spatial control at optical wavelength scales, electron-optical elements can be realized \cite{Freimund, Bucksbaum}. However, this optical control requires light intensities of $10^{14} \,\textrm{W/m$^2$}$. In this paper we report on an optical electron switch that makes use of a small surface and a low power laser. Although some material is placed in the vicinity of the electron beam, no electrical feedthroughs are needed. Moreover, the required laser intensity is reduced by ten orders of magnitude as compared to techniques based on the direct interaction between laser light and electrons.  

In this letter, it is shown that an electron beam that passes by a surface deflects when the surface is illuminated by a low-power continuous-wave diode laser. While searching for a nano-scale related effect at grazing incidence, a significant and unexpected beam deflection was observed. Deflection angles reached value of up-to $1.2 \,\textrm{mrad}$. At a distance of $20 \,\textrm{cm}$ downstream from the interaction region, this translates to a beam displacement of $240 \,\mu\textrm{m}$. A beam-stop can placed in the deflected electron beam, so that chopping the laser light results in complete switching the electron beam on and off. A maximum switching rate of $10^5 \,\textrm{Hz}$ is established. Such an optically controlled electron switch may find applications in electron lithography \cite{MAPPER}, coherent beam splitting or provide an alternative route to STM-based techniques that probe optically induced near-fields \cite{Becker, Elsaesser}. 

\begin{figure}[b]
	\centering
	\scalebox{0.6}{\includegraphics{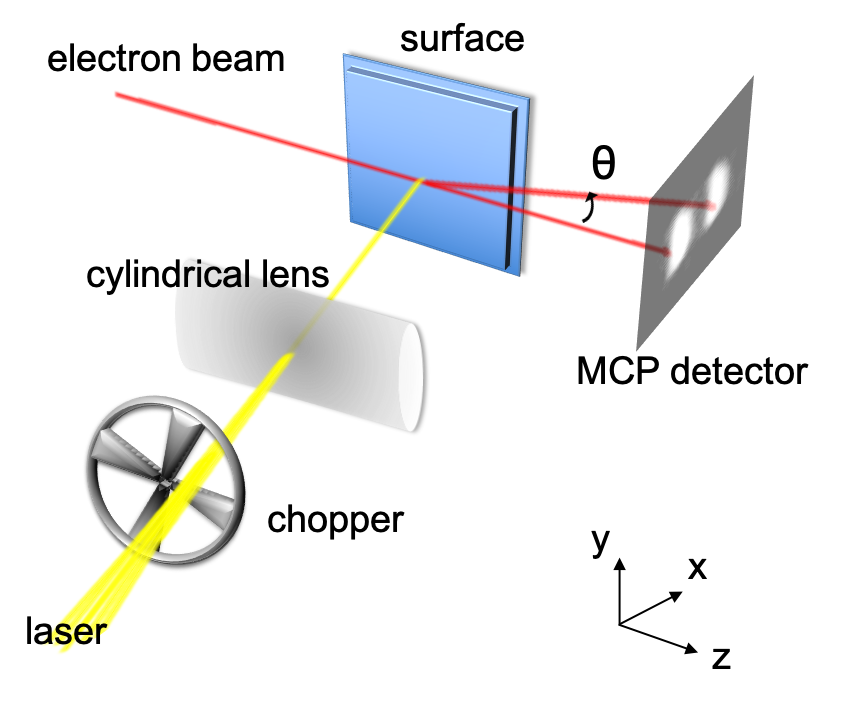}}
	\caption{Setup of the low-power optical electron switch. An electron beam passing close to a surface is deflected by an angle $\theta$ when the surface is illuminated with a laser beam. The illumination is turned on and off with a mechanical chopper. (For a detailed descriptions see text.)}
	\label{switch_setup}
\end{figure}

\begin{figure}[t]
	\centering
	\scalebox{0.5}{\includegraphics{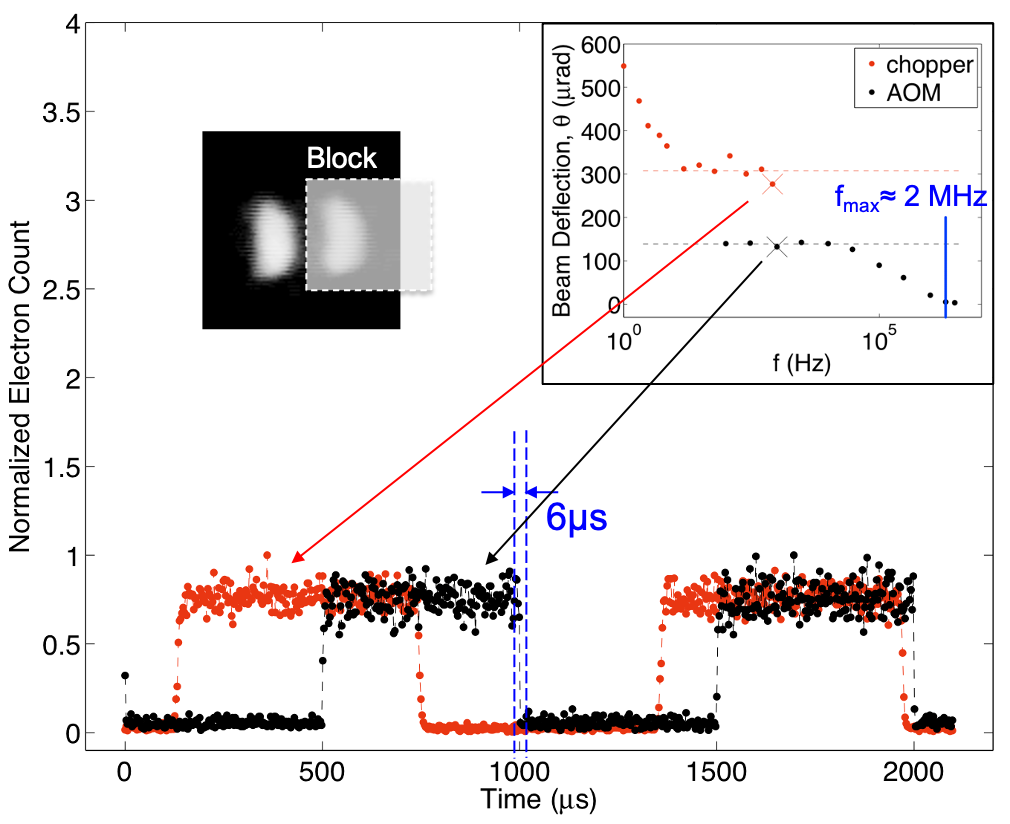}}
	\caption{Electron counts as a function of time as the laser is switching on and off. Both chopper data (red dots) at $818 \,\textrm{Hz}$ and AOM data (black dots) at $1000 \,\textrm{Hz}$ are shown. Top-left inset: A time-averaged image shows the initial and deflected electron beam. A semi-transparent rectangle is added to depict a movable electron beam-stop. Top-right inset: The deflection magnitude $\theta$ is plotted as a function of the chopping frequency $f$. The estimated maximum chopping frequency according our heuristic model, $f_{max} \simeq 2 \,\textrm{MHz}$, is also drawn (blue line) for comparison. The red dots are data collected with a mechanical chopper and the black dots with a AOM.}
	\label{chopping}
\end{figure}

A schematic of the experimental setup is shown in \Fig{switch_setup}. In our experiment, the electron beam is emitted from a thermionic source with a beam energy of $3.98 \,\textrm{keV}$. After passing through two collimation slits of width $5 \,\mu\textrm{m}$ and $2 \,\mu\textrm{m}$ and separation $24 \,\textrm{cm}$, the beam divergence is reduced to $29 \,\mu\textrm{rad}$. At $6 \,\textrm{cm}$ after the second collimation slit, a surface is placed parallel to the beam path. Three different surfaces were tested. The first is a metallic-coated surface with nano-scale grooves \cite{Savas1, Savas2}. The other two are a flat amorphous aluminum (with aluminum oxide on surface) and a silicon nitride surface. All three surfaces resulted in electron beam deflection. 

Continuous-wave diode lasers with maximum powers of $1 \,\textrm{mW}$, $40 \,\textrm{mW}$, and $5 \,\textrm{mW}$ and wavelengths of $532 \,\textrm{nm}$, $685 \,\textrm{nm}$, or $800 \,\textrm{nm}$, respectively, were focused by a cylindrical lens onto the first surface. The other two surfaces were tested with $800 \,\textrm{nm}$ light. The height of the laser beam and electron beam were matched by using an edge of the surface structure to block part of these beams. The focal distance is $25 \,\textrm{cm}$, and the focused laser beam waist was about $280 \,\mu\textrm{m} \times 1 \,\textrm{mm}$ (FWHM). The waist of the light beam was determined by scanning the intensity profile in situ with a surface edge. A $10 \,\mu\textrm{m}$ wide electron beam passes at a distance of nominally $20 \,\mu\textrm{m}$ from the vertically mounted metallic surface. Micrometer stages were used to control the horizontal angle (in the $xz$-plane) as well as the vertical and horizontal travel of the surface. Downstream from the metallic surface, the electron beam passes through a parallel plate electrical deflector that aligns the beam with an electrostatic quadrupole lens. This lens magnifies the electron beam image in the horizontal direction by a factor of $65$. A chevron multi-channel plate (MCP) detector is placed $26 \,\textrm{cm}$ downstream from the surface. A phosphorescent screen follows the multichannel plates and a camera is used to record the beam profile. Amplifiers and discriminators are used in conjunction with a data acquisition board to record the electron counts as a function of time. Gaussian fits of the beam profiles are used to find the center positions and the deflection angles. 

The vacuum pressure is about $1.5 \times {10}^{-7} \,\textrm{Torr}$. By chopping the laser, the electron beam image on the MCP detector switches between two positions. The time-averaged image displays two nearly identical electron beam images that are horizontally displaced from each other (\Fig{chopping}, top-left inset). An electron beam-stop, depicted in the top-left inset of \Fig{chopping} as a semi-transparent rectangle, is added. The electron counts are recorded as a function of time (\Fig{chopping}). The dynamical response of the effect and also the finite electron beam size will limit the rise and fall time. To explore the limit of the response speed, an $40 \,\textrm{MHz}$ acousto-optical modulator (AOM) was used [IntraAction Corp. AOM-40N]. The amplitude of the acoustic wave was modulated from $1 \,\textrm{Hz}$ to $3 \,\textrm{MHz}$. The inset of \Fig{chopping} shows the scaling of the deflection magnitude of the electron beam with the AOM and the chopping frequency. Overall, the deflection magnitude stays constant for frequencies from $10^2 \,\textrm{Hz}$ to $3\times10^5 \,\textrm{Hz}$. When the chopping frequencies are below $10^2 \,\textrm{Hz}$, the deflection magnitude becomes larger. When the AOM frequency increases above $2\times10^5 \,\textrm{Hz}$, the deflection magnitude decreases to zero.

\begin{figure}[t]
	\centering
	\scalebox{0.45}{\includegraphics{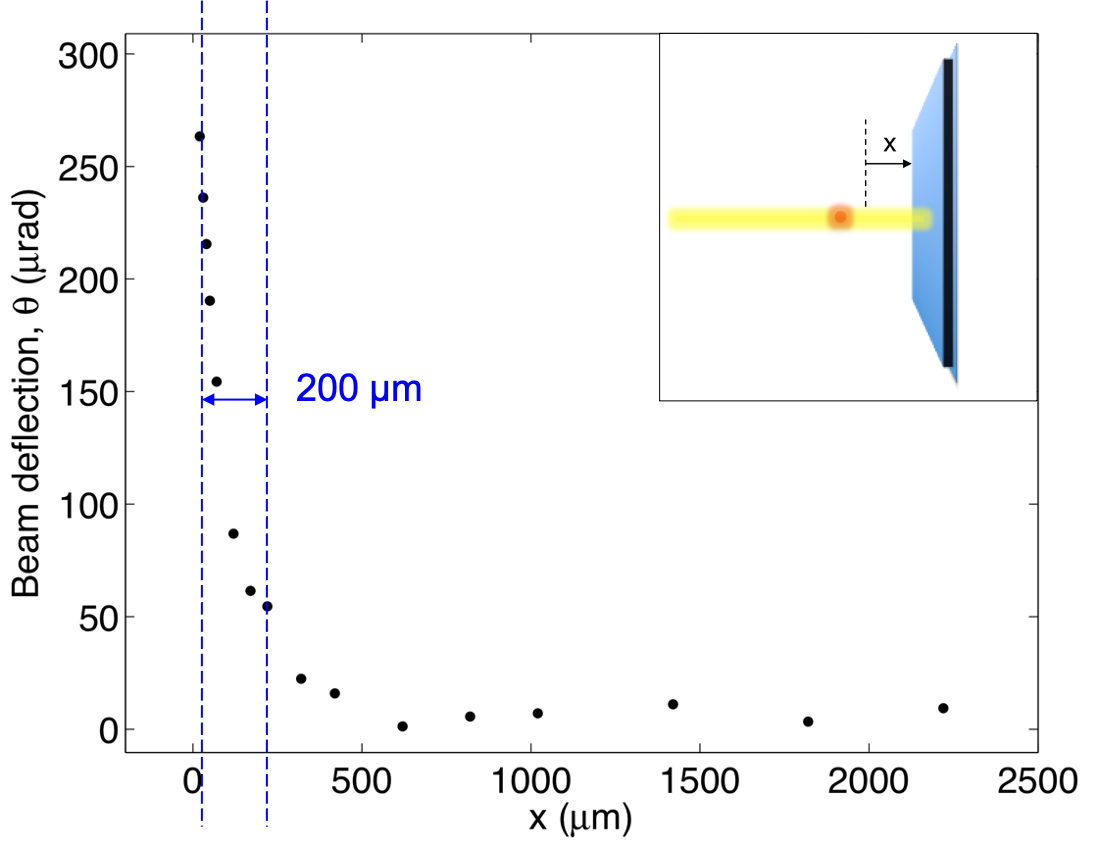}}
	\caption{Distance dependence of the optical electron switch. As the surface is displaced, the distance $x$ between the surface and the electron beam is increased (inset). The optical electron switch turns completely on and off up-to a distance of $200 \,\mu\textrm{m}$.}
	\label{horizontal}
\end{figure}

\begin{figure*}[t]
	\centering
	\scalebox{0.7}{\includegraphics{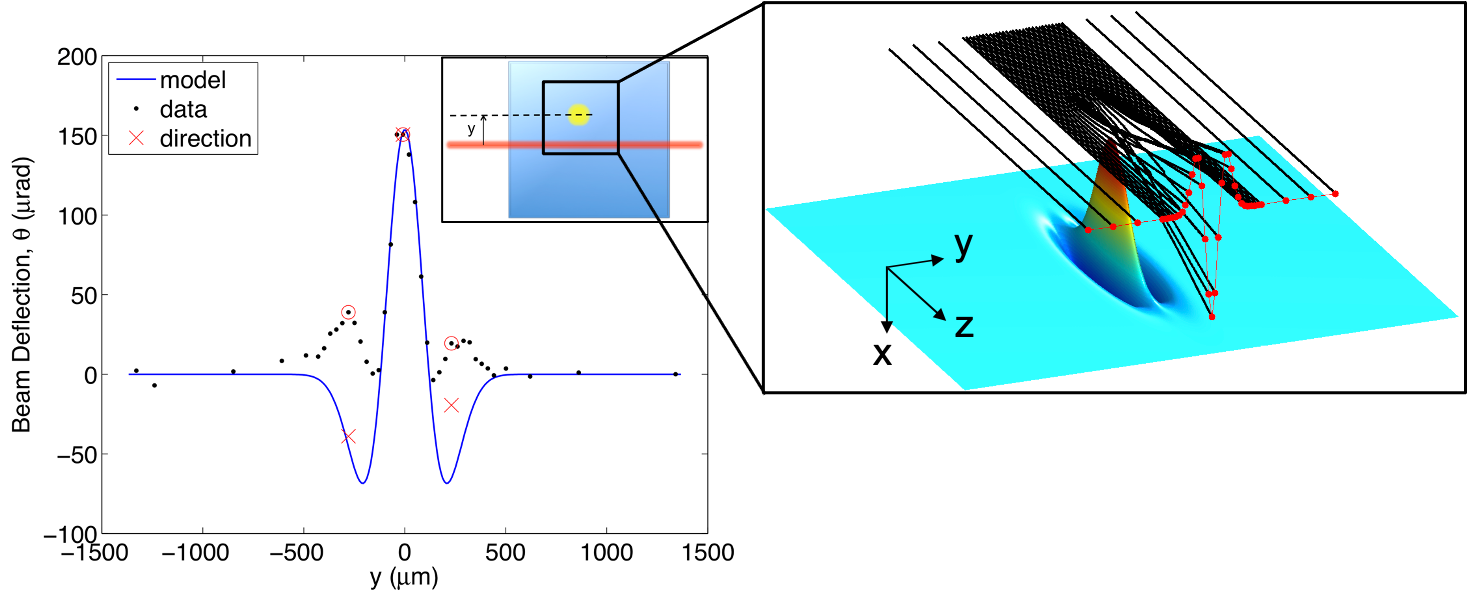}}
	\caption{Beam deflection. Left: The measured deflection magnitude is given as a function of $y$ (black dots). A measurement of the deflection direction is made at three locations (red circles). The values including sign are indicated (red crosses). Reversals of deflection sign may be explained by our heuristic model (blue line) of light-induced surface charge redistribution. Right: A schematics of electron trajectories (black lines) and surface charge density (color-coded) is shown (See text for model description). Red represents positive charge density. Dark blue represents negative charge density. The red dots indicate the final positions of the electron beams. The interaction between the electron beams and the surface charges is attractive in the middle and repulsive at the sides.}
	\label{sign}
\end{figure*}

In \Fig{horizontal}, deflection larger than the beam divergence is observed to a distance of up-to $200 \,\mu\textrm{m}$. The Rayleigh length of the focused laser beam is roughly $5 \,\textrm{cm}$ for an initial beam width of $1 \,\textrm{mm}$ and a the focal length of $25 \,\textrm{cm}$). This is much larger than $200 \,\mu\textrm{m}$, thus the illumination of the surface is unchanged as the surface is moved with respect to the electron beam. This measurement indicates that the deflection originates from the electron-surface interaction rather than the direct electron-laser interaction. As the interaction range is of the order of $200 \,\mu\textrm{m}$, the interacting part of the surface is expected to have a length scale of that order of magnitude. 

When moving the cylindrical lens in the vertical direction, the laser light crosses the electron beam at different heights. The deflection angle shown in \Fig{sign} changes its sign as the light crosses through the electron beam. This was determined by placing the beam-stop in such a way that the electron beam is half-blocked when the laser is off. If the laser light deflects the beam towards the beam-stop, the electron count rate decreases when the light is on. If the laser light deflect the beam away from the beam-stop, the electron count rate increases when the light is on. The magnitude of the deflection is determined by fitting a double Gaussian to the camera image taken with the beam-stop removed. We observed that as the cylindrical lens is moved vertically and the light approaches the electron beam from one end, the electron beam first is deflected away from the surface, then attracted towards the surface, and back to deflected away again. No significant dependence is observed for surface tilt angles or laser polarizations.

Measurements have also been performed on different material surfaces such as uncoated flat silicon-nitride membrane and bulk aluminum. A repulsive deflection of up-to $1.2 \,\textrm{mrad}$ is observed with the silicon-nitride membrane, while at the aluminum surface some small attractive deflection is observed. Given that the deflection effect works with different laser wavelengths at low power, and it can occur at different material surface, we conclude that an optical electron switch based on such a effect is robust. 

In the cases of uncoated silicon-nitride surface, the deflection shows only one sign unlike that observed with the nano-structured metallic-coated surface. This suggests that the deflection mechanism could be complex and involve a host of phenomena including laser heating, plasmon or phonon excitation, and surface-charge redistribution. Nevertheless, a simplistic model is constructed to illuminate some features of our experimental data shown in \Fig{sign}. Focused by the cylindrical lens, the laser intensity profile on the metallic-coated surface can be approximated with an elliptical Gaussian, 
\begin{equation}
	I(y,z) = I_{0} \times exp{\left[ -\left (\frac{y}{\Delta y}\right)^2 -\left (\frac{z}{\Delta z}\right)^2 \right]}.
\end{equation}
where $\Delta y = 170 \,\mu\textrm{m}$ and $\Delta z = 0.6 \,\textrm{mm}$ (corresponding to FWHM of $280 \,\mu\textrm{m} \times 1 \,\textrm{mm}$). The maximum intensity is $I_{0} = P_{0}/(\pi \Delta y \Delta z) = 1.6 \times 10^{4} \,\textrm{W/m$^2$}$ and the laser wavelength is $\lambda = 800 \,\textrm{nm}$. The intensity gradient of the laser light can exert a pondermotive force\footnote{
When a light wave propagates in the solid, the phase relationship between the electric field and the magnetic field is a complex function of the material properties. For a simplistic model, here we assume that the electric field and the magnetic field are in phase.} on the electrons in a thin surface layer, 
\begin{equation}
	\vect{F}_{p} = -\frac{e^2\lambda^2}{8\pi^2 m_{e} c^3 \eps}\vect{\nabla}I.
\end{equation}
If we assume a linear restoring force for the electron,
\begin{equation}\label{harmonic}
¥	\vect{F}_{r} = -\alpha\vect{d},
\end{equation}
where $\alpha$ is a fitting parameter and $\vect{d}$ is the displacement, the induced volume dipole moment can be determined,
\begin{equation}
¥	\vect{P} = -n_{0}e\vect{d} = \frac{1}{\alpha}\frac{n_{0}e^3\lambda^2}{8\pi^{2}m_{e}c^3\epsilon_{0}}\vect{\nabla} I,
\end{equation}
where $n_{0} = 5.9\times10^{28} \,\textrm{m$^{-3}$}$ is the free electron density of gold \cite{Ashcroft}. The volume charge distribution $\rho_{\net}$ is calculated according to $\rho_{\net} = -\nabla\cdot\vect{P}$. Assuming that the pondermotive force is effective through a depth of $\delta_{\eff}=1 \,\textrm{nm}$ into the metal, the effective surface charge distribution on the metallic-coated surface can be obtained,
\begin{equation}
¥	\sigma_{\eff} = \rho_{\net} \delta_{\eff} = - \frac{1}{\alpha} \frac{n_{0}e^3\lambda^2\delta_{\eff}}{8\pi^2m_{e}c^3\epsilon_{0}}\vect{\nabla}^2 I
\end{equation}
The distance between the free electron beam and the surface is $20 \,\mu\textrm{m}$, which is much smaller than the length scale of the surface charge distribution. Thus, close to the surface the free electron beam may experience a electric field approximated by $\vect{E} \simeq \sigma_{\eff}/2\eps (-\hat{\vect{x}})$. Assuming that the velocity is constant in the $z$-direction because of the high kinetic energy $K_{0} = 3.98 \,\textrm{keV}$ in the incoming $z$-direction, the deflection angle of the electron beam along the $x$-axis is estimated by 
\begin{equation}
	\theta = \frac{\Delta v_{x}}{v_{0}} = \frac{e}{4\epsilon_{0}K_{0}} \infint \sigma_{\eff} \,dz. 
\end{equation}
After integration, the above equation becomes
\begin{equation}
¥	\theta = \theta_{0} \left[1-2\left( \frac{y}{\Delta y} \right)^2 \right] e^{ -\left( y/ \Delta y \right)^2 }, 
\end{equation}
where
\begin{equation}
	\begin{split}
	¥	 \theta_{0} &\equiv \frac{\sqrt{\pi}eE_{0}\Delta z}{K_{0}},	\\
		E_{0} &\equiv \frac{\sigma_{0}}{2\epsilon_{0}} ,	\\
		\sigma_{0} &\equiv \frac{1}{\alpha} \frac{n_{0}e^3\lambda^2\delta_{\eff}I_{0}}{8\pi^3\epsilon_{0}m_{e}c^3\Delta y^2}. 	
	\end{split}
\end{equation}
The result of this simplistic model is compared with the experimental data in \Fig{sign}. The fitting parameter is determined to be $\alpha \simeq 1.52\times 10^{-16} \,\textrm{N/m}$. The linear restoring force (\Eq{harmonic}) produces a harmonic motion with fundamental frequency $\omega_{0} = \sqrt{\alpha/m_{e}}$. As a damped harmonic oscillator, the frequency response of the electron switch as shown in the inset of \Fig{chopping} is limited to $f_{max} = \omega_{0}/2\pi \simeq 2 \,\textrm{MHz}$. 

Despite some qualitative agreements, this crude model does not explain many details, such as the physical origin of linear restoring force (\Eq{harmonic}), the increase of the deflection magnitude at very low frequency (\Fig{chopping}), the asymmetric side-peak heights (\Fig{sign}), and the fact that sign reversal of deflection direction is only present on the nano-structured metallic-coated surface but not on the silicon nitride surface. This heuristic model serves to draw attention to these features of our experimental data. 

In summary, when a material surface is placed near an electron beam, a deflection of the electron beam occurs as the surface is illuminated by a low-power laser. Thus, the combination of a material surface, a low-power laser, and a chopping device can make a low-power optical electron switch. Such an optical electron switch may be used for electron beam control in electron lithography and in electron microscopy. 

The qualitative agreement between our model and the experimental data may be fortuitous, but it suggests that the deflection mechanism is consistent with a surface-charge redistribution that is driven by a mechanism that depends on the intensity gradient of the laser light. 

An implication of this work is that instead of using one laser beam for the optical electron switch, one can use multiple laser beams to form spatial-temporal controlled structures on a material surface. The near field of the surface charge may mimic the pattern of the light, and electron matter waves could be coherently controlled in this manner analogous to the Kapitza-Dirac effect or temporal lensing \cite{Batelaan, Baum}, but without the need for high laser intensity. Finally, we speculate that the combination of laser pulses and nano-fabricated structures will make femtosecond manipulation of free electrons accessible at low intensities \cite{Becker, Zewail, Muskens}.  

We gratefully acknowledge the funding support from NSF Grant No.~0969506.

\end{document}